\documentclass[12pt]{iopart}

\usepackage[english]{babel}
\usepackage{color}

\usepackage{amscd}

\tiny
\def\d{{\rm d}}

\def\cI{{\cal I}}

\def\cIL{{\cal I}_{\rm L}}

\def\hcI{\hat{\cal I}}

\def\hcIL{\hat{\cal I}_{\rm L}}

\def\hcIDM{\hat{\cal I}_{\rm DM}}

\def\hcIGDM{\hat{\cal I}_{\rm GDM}}

\usepackage{iopams}  
\begin{document}

\title[On the Lewis--Riesenfeld (Dodonov--Man'ko) invariant method]
{On the Lewis--Riesenfeld (Dodonov--Man'ko) invariant method}

\author[J. Guerrero and F.F. L\'opez-Ruiz] 
{Julio Guerrero and Francisco F. L\'opez-Ruiz}

\address{}
\ead{juguerre@um.es, 	paco.lopezruiz@uca.es}

\begin{abstract}

We revise the Lewis--Riesenfeld invariant method for solving the quantum time-dependent harmonic oscillator 
in light of the  Quantum Arnold Transformation previously 
introduced and its recent generalization to the 
Quantum Arnold--Ermakov--Pinney Transformation. We prove that both methods are equivalent and
show the advantages of the Quantum Arnold--Ermakov--Pinney transformation over the 
Lewis--Riesenfeld invariant method. We show that, in the quantum time-dependent and damped harmonic oscillator, the invariant proposed
by Dodonov \& Man'ko is more suitable and provide some examples to illustrate it, focusing on the damped case.
\end{abstract}


%

\section{Introduction}

The Lewis--Riesenfeld  invariant method \cite{Lewis,LewisRiesenfeld} is a technique that
allows to obtain a complete set of solutions of the Schr\"odinger equation for a time-dependent 
harmonic oscillator in terms of the eigenstates of a quadratic invariant. This quadratic invariant,
the Lewis invariant, is built using an auxiliary variable that satisfies the
Ermakov equation \cite{Ermakov,Ermakov-trad,Milne,Pinney}. 

The Quantum Arnold--Ermakov--Pinney Transformation (QAEPT) \cite{QAEPT} is a unitary transformation that 
maps solutions of a Generalized Caldirola--Kanai  \cite{Caldirola,Kanai,QAT} 
Schr\"odinger equation into solutions of another Generalized Caldirola--Kanai Schr\"odinger equation. In particular, one of 
the systems can be the standard harmonic oscillator and the other a time-dependent 
harmonic oscillator, and in this case we shall show that  the Lewis--Riesenfeld invariant method is recovered.

The idea of using invariants to solve equations is rather old, going back to S. Lie (1883)
\cite{Lie} who showed that a second order differential equation has the maximal group of
symmetries 
if the differential equation 
is up to third order in the derivative, and the coefficients satisfy certain relations \cite{Mahomed-Qadir,Aminova}. 

V.P. Ermakov  (1880) \cite{Ermakov,Ermakov-trad} showed that the general solution of the non-linear 
equation 
\begin{equation}
\ddot b + \omega^2(t) b = \frac{\omega_0^2}{b^3} \,,
\label{EMP}
\end{equation}
where $\omega_0$ is an arbitrary constant, can be obtained from two 
independent  solutions $y_1, y_2$  of the corresponding linear
equation:
\begin{equation}
 \ddot y + \omega^2(t) y = 0
\end{equation}
by: 
\begin{equation}
 b^2=c_1 y_1^2+ c_2 y_2^2 + 2 c_3 y_1 y_2, 
\end{equation}
with $c_1c_2-c_3^2=\omega_0^2$. Similar results were derived independently by W.E. Milne (1930)  \cite{Milne} 
and E. Pinney (1950) \cite{Pinney}.

H.R. Lewis (1967) \cite{Lewis} obtained a classical and quantum quadratic invariant for 
a time-dependent 
harmonic oscillator, of the form: 
\begin{equation}
 \cIL = \frac{1}{2m} (b p - m \dot b x)^2 + \frac{1}{2}m\omega_0^2 \frac{x^2}{b^2}\,,
 \label{EL-invariant}
\end{equation}
where, again, $\omega_0$ is an arbitrary  constant with dimension of frequency, in terms of an auxiliary dimensionless function $b(t)$
satisfying the Ermakov equation (\ref{EMP}). 
Note that classically $x$ satisfies the equation of motion 
\begin{equation}
 \ddot x + \omega^2(t) x = 0 \,.\label{HO}
\end{equation}
The pair of equations (\ref{EMP}) and (\ref{HO}) is denoted an Ermakov system. They are uncoupled 
(given $\omega(t)$, they can be solved independently  for $x$ and $b$),
although it has been generalized to coupled equations and to higher dimensions 
\cite{ErmakovSystems1,ErmakovSystems2}.

The reader might wonder about the comparison of $b$ (dimensionless), satisfying (\ref{EMP}) and 
providing the invariant (\ref{EL-invariant}), and a function $\rho$ satisfying the usual 
Ermakov--Pinney equation  $\ddot \rho + \omega(t)^2 \rho = \frac{1}{\rho^3}$, which has the 
dimensions of the square root of time (see e.g. \cite{Lewis}). The relation between $b$ and $\rho$ is 
simply $b=\sqrt{\omega_0} \rho$. Also, the relation between $\cIL$ and the invariant $I$ in 
\cite{Lewis} is simply $\cIL = \omega_0 I$, i.e. $\cIL$ has dimension of energy meanwhile $I$ has dimension
of action. The arbitrariness in the choice of $b$ and $\cIL$ 
was already noted in \cite{Lewis,LewisRiesenfeld} and we make it explicit for later convenience (see also 
\cite{InverseEngineer}, where the authors use the same convention).

Lewis \& Riesenfeld  (1969) 
\cite{LewisRiesenfeld} used the eigenvectors of the quantum version of this quadratic invariant $\hcIL$ written in terms 
of the auxiliary function $b(t)$ satisfying the Ermakov equation to obtain solutions of 
the  Schr\"odinger equation for a time-dependent 
harmonic oscillator. For this purpose, an extra time-dependent
phase $e^{i\int\frac{\omega_0}{b^2}dt}$ had to be added to the eigenfunctions in order to satisfy 
the Schr\"odinger equation. They did not considered damping (or time dependent mass) and
they supposed that the quadratic invariant has discrete spectrum.

V.I. Arnold (1978) \cite{Arnold}, in the context of symmetries of second order 
ordinary differential equations,  introduced the term \textit{straightening} for the
linearization studied by S. Lie and considered the case of Linear Second Order
Differential Equations (LSODE):
\begin{equation}
\ddot{x}+\dot{f}\dot{x} +\omega^{2}x=\Lambda\,,
\label{LSODE}
\end{equation}
where $f$, $\omega$ and $\Lambda$ are time-dependent functions,
giving explicitly the transformation for this case:
\begin{equation}
\begin{array}{rccl}
A :& \mathbb{R}\times T&\rightarrow &\mathbb{R}\times{\cal T}\\
&(x,t)&\mapsto &(\kappa,\tau)
\end{array}
\label{CAT}
\end{equation}
with
\begin{equation}
\tau = \frac{u_{1}(t)}{u_{2}(t)}\,, \qquad
\kappa =  \frac{x - u_p(t)}{u_{2}(t)}\,,
\end{equation}
where $T$ and ${\cal T}$ are, in general, open intervals; 
$u_1$ and $u_2$ are independent solutions of the homogeneous LSODE, $u_p$ is a particular 
solution of the inhomogeneous LSODE, and 
$W(t)=\dot{u}_1u_2-u_1\dot{u}_2=e^{-f}$ is the Wronskian of the two solutions\footnote{If $x$ represents position
and $t$ represents time, then $u_1$  has dimension of time,  $u_2$ is dimensionless and $u_p$ has dimension of length.
}.

Under this transformation, the classical equation of motion (\ref{LSODE}) transforms as:
\begin{equation}
\ddot{x}+\dot{f}\dot{x} +\omega^{2}x=\Lambda  \quad
  \stackrel{A}{\longrightarrow}\quad \frac{W}{u_2^3}\,\, \ddot \kappa =0\,.
\label{CAT_app}
\end{equation}
Thus, the Arnold transformation maps patches of solutions of the LSODE system into 
patches of free particle trajectories.

For convenience, we shall impose the canonicity conditions (see \cite{QAT}):
\begin{equation}
 u_1(0)=\dot{u}_2(0)=u_p(0)=\dot{u}_p(0)=0\,, \quad \dot{u}_1(0)=u_2(0)=1\,.\label{canonicity}
\end{equation}
These conditions play a crucial role in the physical interpretation of quantities mapped
from one system into the other through the Arnold transformation. More precisely, if ${\kappa}(\tau)$ and ${\pi}(\tau)$ are the 
conserved position and momentum for the free particle (verifying that ${\kappa}(0)=\kappa$ and 
${\pi}(0)=\pi\equiv m\dot\kappa$), then the transformed quantities through the Arnold transformation are 
the conserved position $x(t)$ and momentum $p(t)$ in the LSODE system (verifying $x(0)=x$ and 
$p(0)=p\equiv m\dot x$).

Dodonov \& Man'ko (1979) \cite{DodonovManko,DodonovManko-libro} (and Malkin--Man'ko--Trifonov \cite{MalkinMankoTrifonov}
(1969) without considering 
damping) computed the coherent states for the Generalized Caldirola--Kanai  model 
(the quantum version of a general LSODE system), whose 
Hamiltonian is
\begin{equation}
 \hat{H}_{GCK}=\frac{\hat{p}^{2}}{2m}e^{-f}+\bigl(\frac{1}{2}m\omega^{2}\hat{x}^{2} -m\Lambda \hat{x}
\bigr)e^{f}\,,
\label{GCK-Hamiltonian}
\end{equation}
using first-order invariants as annihilation and creation operators. The number operator associated with 
these annihilation and creation operators is a quadratic invariant that will be denoted the Dodonov--Man'ko invariant $\hcIDM$.
 Later, other authors have used 
first-order invariants to solve time-dependent problems \cite{Pedrosa2}.

Hartley \& Ray (1981) \cite{HartleyRay} and Lewis \& Leach (1982) \cite{LewisLeach}
generalized the construction of the Lewis invariant to some non-linear systems. 


Pedrosa (1987) \cite{Pedrosa} constructed the Lewis invariant for the Ermakov equation with 
a damping term using canonical transformations.

V. Aldaya et al. (2011) \cite{QAT} extended to the quantum case the Arnold transformation
and denoted it the Quantum Arnold Transformation (QAT): 
\begin{eqnarray}\eqalign{
  \hat {A}:& \quad {\mathcal H_t} \;\; \longrightarrow
\quad {\mathcal H^G_\tau} \\
    & {\phi}(x,t)  \longmapsto \;
{\varphi}(\kappa,\tau) = 
        \hat A \left( {\phi}(x,t) \right) \\
     & \qquad \qquad = 
       A^* \left( \sqrt{u_{2}(t)}\,e^{-\frac{i}{2}\frac{m}{\hbar}
                 \frac{1}{W(t)}\frac{\dot{u}_{2}(t)}{u_{2}(t)} {x}^{2}}
{\phi}(x,t) \right) \,.} \label{QAT}
\end{eqnarray}
Here $A^*$ is 
defined as $A^*(f(x,t))=f(A^{-1}(\kappa,\tau))$,  
${\mathcal H_t}$ is the Hilbert space of solutions of the Generalized Caldirola--Kanai Schr\"odinger equation at time $t$,
and ${\mathcal H^G_\tau}$ is the Hilbert space of solutions of the Schr\"odinger equation
for the free Galilean particle at time $\tau$, where $t$ and $\tau$ are related by the Arnold Transformation.
Note that the QAT transforms solutions of the \textit{time-dependent} Schr\"odinger equation of the Generalized Caldirola--Kanai system
into free-particle wave functions, and that this is achieved by applying the Arnold transformation
\textit{together with} multiplying the wave function by a suitable phase and rescaling factor. These factors
also renders the QAT unitary \cite{QAT,QAEPT}.

Some applications of the QAT were given in \cite{HarmonicStates}, where states from the harmonic oscillator were mapped into the free particle giving rise to Hermite--Gauss and
Laguerre--Gauss wave packets; in \cite{CQAT,NuovoCimento}, where processes of Release and Recapture of a particle by a harmonic trap were studied using the QAT; and in
\cite{UnfoldedQAT}, where the QAT, which is a local diffeomorphism in time, is extended beyond the ``focal'' points, correctly reproducing the change in phase of the wave function 
(Maslov correction, see for instance \cite{Horvathy}).

Casta\~nos, Schuch \& Rosas-Ortiz (2013) \cite{Castanos-etal} constructed
coherent states for different models (time-dependent and non-linear Hamiltonians) 
through complex Riccati equations and found the corresponding Lewis invariants. 

Since its introduction, the Lewis invariant and its associated 
Ermakov equation entered an inflationary scenario with applications in many areas.
One of the most remarkable ones are the applications in Bose--Einstein Condensates (BEC) \cite{Kagan,Castin},
 where a transformation similar to that of Arnold (and known as scaling transformation in this 
context) taking the time-dependent harmonic trap in the Gross--Pitaevskii equation into a 
stationary one is applied. Although they did not use the Lewis invariant, the scaling parameter satisfies
the Ermakov equation (\ref{EMP}). In this context, eq. (\ref{EMP}) also appears in \cite{Garcia-Ripoll_etal}.

Recently, the Lewis--Riesenfeld invariant method has been used to inverse engineer short-cuts
to adiabaticity \cite{InverseEngineer}, to speed up cooling processes and transport in 
electromagnetic traps and BECs, and to manipulate states in wave-guides \cite{Shortcut-waveguides}, where the relation with Generalized Caldirola--Kanai systems has been established \cite{waveguides}. 
The main idea here is to design a Lewis invariant satisfying 
the property of commuting with the Hamiltonian at initial and  final times, and this can be achieved
by building up a function $b$ satisfying certain boundary conditions and then determining, 
through the Ermakov equation, the time-dependent frequency that should be applied  in 
order to take the system from the initial state to the desired final state without affecting to
the population of the levels.

Another recent application of the  Lewis--Riesenfeld invariant method is in mesoscopic 
RLC electric circuits \cite{RLCcirtuits}, where the quantum evolution  (even in the case of
 time-dependent $R(t)$, $L(t)$ and $C(t)$ and source term) is described.

 The content of the paper is as follows. In Sec. \ref{SecELR} we revise the  Lewis--Riesenfeld method and explain it in
 terms of the QAT and the QAEPT, showing that the use of the Dodonov--Man'ko invariant is more appropriate for damped systems.
 In Sec. 3 the examples of the Caldirola--Kanai and the Hermite oscillators are studied in detail.

 \section{The Lewis--Riesenfeld method in light of 
the Quantum Arnold Transformation}
\label{SecELR}


In their original paper Lewis \& Riesenfeld \cite{LewisRiesenfeld} provided a method to obtain a family of exact wave functions for 
the time-dependent 
harmonic oscillator spanning the whole Hilbert space. In a first step the method looks for an invariant,  
Hermitian operator $\hcIL$, a task which can follow the lines of \cite{Lewis}. Imposing the invariance 
condition\footnote{Although we are working in the Sch\"odinger picture of Quantum Mechanics, i.e. wavefunctions
depend explicitly on time and common operators like position $\hat{x}$ and momentum $\hat{p}$ do not depend on time,
other quantum operators may depend explicitly on time. This is precisely the case of the Hamiltonian for
non-conservative systems and in general for invariant operators.}
\begin{equation}
\frac{\d \hcIL}{\d t} \equiv \frac{\partial \hcIL}{\partial t} 
   + \frac{i}{\hbar}[\hat{H}(t),\hcIL] = 0\,,
  \label{invariancia}
\end{equation}
where $\hat{H}(t)$ is the Hamiltonian for the time-dependent 
harmonic oscillator and assuming the most general quadratic invariant,  
they arrived at the quantum version of (\ref{EL-invariant}), where the auxiliary function $b(t)$ satisfies the Ermakov 
equation (\ref{EMP}). In this equation $\omega_0$ is just an arbitrary constant. The possibility exists of giving a 
generic form for the invariant (i.e. quadratic or linear, Hermitian or complex combinations of basic operators 
$\hat{p}$ and $\hat{x}$, etc.) and then solving for the coefficients to fulfill (\ref{invariancia}). 

The second step in the method is realizing that finding eigenfunctions $\phi_s(x,t)$ of $\hcIL$,  $\hcIL \phi_s(x,t) = \lambda_s \phi_s(x,t)$, amounts to finding solutions of the Schr\"odinger equation except for a time-dependent phase, which must be computed. That is, solutions $\psi_s(x,t)$ of the Schr\"odinger equation

\begin{equation}
i\hbar\frac{\partial \psi_s}{\partial t}=-\frac{\hbar^{2}}{2m}
\frac{\partial^{2} \psi_s}{\partial x^{2}}+\frac{1}{2}m\omega^{2}(t)x^{2}
\psi_s \equiv \hat{H}(t) \psi_s
\label{eq_schrodinger_TDHO}
\end{equation}
may be of the form 

\begin{equation}
\psi_s(x,t) = e^{i \alpha_s(t)} \phi_s (x,t)\,,
\end{equation}
where $ \alpha_s(t)$ satisfies

\begin{equation}
\hbar \frac{\d \alpha_s(t)}{\d t} \phi_s (x,t) = \Bigl( i\hbar\frac{\partial}{\partial t}- \hat{H}(t)\Bigr)\phi_s (x,t) \,.
\end{equation}

That is just a nice consequence of the fact that $\hcIL$ applied on a solution of (\ref{eq_schrodinger_TDHO}) is again a solution: 

\begin{equation}
i\hbar\frac{\partial (\hcIL \psi_s)}{\partial t} = \hat{H}(t) (\hcIL \psi_s)\,. 
\end{equation}
The phase can be solved in terms of $b$ to give: 
\begin{equation}
 \alpha_s(t)=-\frac{\lambda_s}{\hbar}\int\frac{dt}{b(t)^2}\,.
\end{equation}

Two observations can be made. First, the way in which the eigenfunctions 
$\phi_s(x,t)$ of the invariant are found are left to the ability of the user of the method. In this respect, some authors have developed a unitary transformation from the Hamiltonian of the simple harmonic oscillator into the invariant \cite{Moya-Cessa_Fernandez-Guasti} (resembling very much the QAEPT, see below). For this purpose, transforming the time-independent Schr\"odinger equation of the simple harmonic oscillator, including the wave functions, would do the trick. 
Second, 
%
although the Lewis-Riesenfeld method can provide all (quadratic) invariants for the time dependent harmonic
oscillator (by taking all possible solutions of the Ermakov equation) it does not provide insight on their
physical interpretation (like their spectra). In our case, we provide a method that allows a neat physical
interpretation of the invariants since each one preserves the same character as in the harmonic oscillator.

In that sense the QAT and its generalization the QAEPT turn out to  be very useful.

\subsection{The Quantum Arnold--Ermakov--Pinney transformation}

The QAEPT is obtained when two different LSODE-systems are related by QATs with the free-particle system as an intermediary, that is, when a QAT and an inverse QAT are composed.  That was shown in \cite{QAEPT}. As the QAT, the QAEPT relies on the symmetry structure of the systems of the Generalized Caldirola--Kanai type (see \cite{QAT}). 


In the QAEPT, it is the full set of invariant operators 
\textit{and} the corresponding eigenstates including the time dependence (with no need of searching for a phase as it is given by the transformation) which is mapped from the simple harmonic oscillator system into a Generalized Caldirola--Kanai system. The interpretation of the eigenstates may be the same on both sides of the mapping. 

Let $A_1$ and $A_2$ denote the Arnold transformations relating the LSODE-system 1 and LSODE-system 2  to the free particle, respectively, then $E=A_1^{-1}A_2$ relates LSODE-system 2 to LSODE-system 1. $E$ can be written as:
\begin{eqnarray} \eqalign{
 E: &\,\mathbb{R}\times T_2\rightarrow \mathbb{R}\times T_1 \\
&(x_2,t_2)\mapsto (x_1,t_1)=E(x_2,t_2)\,.}
\end{eqnarray}

The explicit form of the transformation can be easily computed by composing the two Arnold transformations, resulting in:
\begin{equation}
x_1=\frac{x_2}{b(t_2)}\qquad W_1(t_1)dt_1=\frac{W_2(t_2)}{b(t_2)^2}dt_2\,,
\label{Arnold-Ermakov-Pinney}
\end{equation}
where $b(t_2)=\frac{u^{(2)}_2(t_2)}{u^{(1)}_2(t_1)}$ satisfies the non-linear SODE:
\begin{equation}
 \ddot b + \dot f_2 \dot b + \omega_2^2 b = \frac{W_2^2}{W_1^2}\frac{1}{b^3}
            \left[\omega_1^2+\dot f_1 \frac{\dot u^{(1)}_2}{u^{(1)}_2}(1-b^2\frac{W_1}{W_2})\right]\,,
\label{GEP}
\end{equation}
and where $u^{(j)}_i$ refers to the $i$-th particular solution for system $j$; $W_j$, $\dot f_j$ and $\omega_j$ stand for the 
Wronskian and the LSODE coefficients for system $j$; and the dot means derivation with respect to the corresponding time variable.
If all  $u^{(j)}_i$ satisfy the canonicity conditions (\ref{canonicity}) then $b(t_2)$ satisfies the corresponding 
canonicity conditions
\begin{equation}
 b(0)=1\,,\qquad \dot{b}(0)=0\,. \label{canonicity-b}
\end{equation}

Equation (\ref{GEP}) constitutes a \textit{generalization} of the Ermakov equation. That equation, together with the LSODE of system 2, is a \textit{generalized Ermakov pair} \cite{ErmakovSystems1,ErmakovSystems2}. Also, any (quadratic) conserved quantity, 
which is shared by the two LSODE-systems, constitutes a \textit{generalized Lewis invariant}. 
Equation (\ref{GEP}) actually \textit{defines} a generalized Arnold transformation, to be named (classical) \textit{Arnold--Ermakov--Pinney transformation}, which transforms solutions of the LSODE 1 into solutions of the LSODE 2. 

The quantum version of the Arnold--Ermakov--Pinney transformation, $\hat E$, 
can be obtained by computing the composition of a QAT and an inverse QAT  to give:
%
\begin{eqnarray}\eqalign{
  \hat {E}:& \quad {\mathcal H^{(2)}_{t_2}} \;\; \longrightarrow
\quad {\mathcal H^{(1)}_{t_1}} \\
    & {\phi}(x_2,t_2)  \longmapsto \;
{\varphi}(x_1,t_1) = 
        \hat E\left( {\phi}(x_2,t_2) \right) \\
          &\qquad\qquad= E^* \left( \sqrt{b(t_2)}\,e^{-\frac{i}{2}\frac{m}{\hbar}
                 \frac{1}{W_2(t_2)}\frac{\dot b(t_2)}{b(t_2)} {x}^{2}_2}{\phi}(x_2,t_2) \right) \,.}
\label{AEP}                
\end{eqnarray}

The Quantum Arnold--Ermakov--Pinney transformation (QAEPT) maps solutions of a Generalized Caldirola--Kanai Schr\"odinger equation into solutions of a different, auxiliary Generalized Caldirola--Kanai Schr\"odinger equation, and by construction
it is also a unitary transformation. The auxiliary system might be, 
in particular,  the one corresponding to a harmonic oscillator with frequency $\omega_1(t)=\omega_0$ and $\dot{f}_1=0$.
In this case eqs. (\ref{Arnold-Ermakov-Pinney}) and (\ref{GEP}) reduce to: 
\begin{equation}
x_1=\frac{x_2}{b(t_2)}  \,,\qquad  t_1=\int^{t_1}_0\frac{W_2(t)}{b(t)^2}dt\,,
\label{Arnold-Ermakov-Pinney2}
\end{equation}
and
\begin{equation}
 \ddot b + \dot f_2 \dot b + \omega_2^2 b = \frac{W_2^2\omega_0^2}{b^3}\,.
\label{GEP2}         
\end{equation}


\subsection{Ermakov System and interpretation of the Lewis Invariant}

Consider the particular case where LSODE-system 1 is a harmonic oscillator ($\omega_1(t_1)\equiv \omega_0$ and $\dot f_1=0$), which can be described by the Hamiltonian
\begin{equation}
H_{HO}=\frac{p_1^{2}}{2m}+\frac{1}{2}m\omega_0^{2}x_1^{2}\,,
\label{HOHam}
\end{equation}
and LSODE-system 2 is a time-dependent harmonic oscillator with frequency $\omega_2 (t_2) \equiv \omega(t)$ and $\dot f_2=0$, with Hamiltonian $H(t)$ given by (\ref{eq_schrodinger_TDHO}).
Then, expressions (\ref{GEP}) and (\ref{GEP2})  simplify to (\ref{EMP}).
Obviously, for $\omega_0=0$ the Arnold--Ermakov--Pinney transformation reduces to the ordinary Arnold transformation, i.e.  $E=A$.

Now, note that LSODE 1 Hamiltonian, $H_{HO}$, is conserved, and that it is so on both sides of the transformation $E$, given by (see (\ref{Arnold-Ermakov-Pinney})): 
\begin{equation}
x_1=\frac{x}{b}\,,\quad t_1=\int\frac{1}{b^2}dt = \frac{1}{\omega_0}\arctan\omega_0\tau\,,
\label{AEPT}
\end{equation}
where $\tau$ denotes the (common) time in the free particle given by the Arnold transformations $A_1$ and $A_2$. 
Also, $b(t)=u_2^{(2)}\sqrt{1+\omega_0^2\tau^2}=\sqrt{(u_2^{(2)})^2+\omega_0^2(u_1^{(2)})^2}$ satisfies the Ermakov equation
(\ref{EMP}) together with the canonicity conditions (\ref{canonicity-b}).

It should be stressed that  $b(t)$ never vanishes, otherwise the
Wronskian $W_2(t)$ of the two independent solutions would also vanish. And since in the quantum case the time
$t_1$ appears in the form $e^{-i\omega_0 t_1}=e^{-i\arctan(\omega_0\tau)}$, this 
expression is well-defined for all times (even in the case where $\tau$ has singularities).
This means that the QAEPT transformation is well defined for all times 
(i.e. $E:\mathbb{R}\times\mathbb{R}\rightarrow\mathbb{R}\times\mathbb{R}$). 
This is an important advantage with respect to the QAT,
that was defined only locally in time.

Computing the momentum $p_1 = m \dot x_1 = m \frac{d x_1}{d t_1}= m \frac{d t}{d t_1}
\frac{d}{d t} (\frac{x}{b}) = m (\dot x b-\dot b x)$, we can write $H_{HO}$ in variables corresponding to system 2:
\begin{equation}
H_{HO}= \frac{1}{2m}(p b - m \dot b x)^2 + \frac{1}{2}m \omega_0^2 (\frac{x}{b})^2  \equiv \cIL\,.
\label{HOHamtrans}
\end{equation}
That is easily recognized as the usual Lewis invariant $\cIL$. Thus, we have found a way to characterize it through the 
Arnold--Ermakov--Pinney Transformation: $\cIL$ corresponds to the conserved quantity $H_{HO}$ imported from the simple harmonic oscillator, which is used as an auxiliary system. Because the auxiliary system is arbitrary, $\cIL$ is conserved for any $\omega_0$, provided (\ref{EMP}) is satisfied. 
Note that, in order to establish the identification $H_{HO}\equiv \cIL$ it is essential to impose the canonicity
conditions (\ref{canonicity-b}).

Using the explicit form of the inverse $\hat E^{-1}$ of (\ref{AEP})  in this case, it is straightforward to arrive at solutions $\phi(x,t)$ of the Schr\"odinger equation of the time-dependent harmonic oscillator in terms of solutions of the Schr\"odinger equation for the simple harmonic oscillator $\varphi(x_1,t_1)$: 
\begin{equation}
\phi(x,t)=\frac{1}{\sqrt{b}}e^{\frac{i}{2}\frac{m}{\hbar}
                 \frac{\dot b}{b} {x}^{2}}{\varphi}\Big(\frac{x}{b},\int \frac{1}{b^2} dt\Big)\,,
                \label{TDHO_sol}
\end{equation}
where $b$ is any solution of (\ref{EMP}) satisfying the canonicity conditions (\ref{canonicity-b}). Note that, if $\varphi(x_1,t_1)$ is chosen to be, for instance, an eigenfunction of the quantum operator corresponding to (\ref{HOHam}), $\hat{H}_{HO}$, then the transformed wave function $\phi(x,t)$ is an  eigenfunction of the quantum operator $\hcIL$ corresponding to the invariant (\ref{HOHamtrans}) (the explicit form of such operators is easily obtained from their classical counterpart by the canonical quantization prescription). That shows that $\hcIL$ has discrete spectrum. 

In the wave functions (\ref{TDHO_sol}), two phases can be distinguished: the one corresponding to the transformation itself, explicit in (\ref{TDHO_sol}), and the phase mapped from $e^{-i (n+\frac{1}{2}) \omega_0 t_1}$ (which is the only time dependence for stationary states in the harmonic oscillator) into $e^{-i (n+\frac{1}{2}) \omega_0 \int\frac{dt}{b(t)^2}}$. The latter accounts for the phase of the Lewis--Riesenfeld method. The former accounts for the phase (and the factor) which appears, for instance, in \cite{Moya-Cessa_Fernandez-Guasti}.

Regarding the canonicity conditions (\ref{canonicity-b}), they play an important role in short-cuts to adiabaticity processes for time-dependent harmonic oscillators, 
see \cite{InverseEngineer}, since they imply that $\hcIL$  commute with the Hamiltonian at the initial time $t=0$. If we further
impose $\ddot b=0$, then $\omega_0=\omega_2(0)$ holds, and the Hamiltonian at
$t=0$ will coincide with the invariant $\hcIL$ at $t=0$. In the following, we shall assume
that the Lewis invariant $\hcIL$ verifies these conditions.

The same process can be repeated for any other operator representing an invariant in the simple harmonic oscillator (LSODE 1), showing the usefulness of the QAEPT to perform quick computations. 

In conclusion, it is the full set of invariant operators \textit{and} the corresponding eigenstates (with no need of searching a phase) what is mapped from the simple system into the Generalized Caldirola--Kanai system through the QAEPT. Also, a word of caution is in order: the Hamiltonian operator of one system is not mapped into the  Hamiltonian operator of the other system, which may not be invariant itself. 


\subsection{The Lewis--Riesenfeld (Dodonov-Man'ko) invariant method for the Generalized Caldirola--Kanai oscillator through the QAEPT}
\label{Invariant-GCK}

In a Generalized Caldirola--Kanai system, the easiest way to find eigenstates of an  invariant operator
and its eigenfunctions as solutions of the Generalized Caldirola--Kanai Schr\"odinger equation, is to focus on an auxiliary system (the harmonic oscillator in the previous subsection) with its Hamiltonian being the invariant operator and perform the QATs or QAEPT necessary to map the Schr\"odinger equation of such auxiliary system into the Generalized Caldirola--Kanai Schr\"odinger equation. 
In this process, Hamiltonian is not mapped into Hamiltonian, but conserved operators into conserved operators are. That procedure takes advantage of the fact that the eigenstates of the harmonic oscillator Hamiltonian have a very simple time dependence, as has just been noticed in the previous Subsection. 

Let us now describe a different way of constructing an invariant. The idea is to consider
any linear combination of quadratic invariants
in such a way that its
eigenfunctions solve the Generalized Caldirola--Kanai Schr\"odinger equation. 
%
The most general invariant can be written in the form \cite{QAT}:
\begin{equation}
  \hcI = \frac{1}{2m} \hat P^2 + \frac{1}{2} m \tilde \omega^2 \hat X^2 +
\frac{\tilde \gamma}{2} \;\frac{ \hat{X}\hat{P}+\hat{P}\hat{X}}{2}\,. 
\label{DMinvariant}
\end{equation}
where $\tilde{\omega}$ and $\tilde{\gamma}$ are arbitrary real numbers and
$\hat{X},\hat{P}$ are conserved position and momentum operators satisfying that at $t=0$
coincide with the usual $\hat{x},\hat{p}$, namely \cite{QAT}:
\begin{equation}
\hat P = -i \hbar u_2 \frac{\partial}{\partial x} - m x \frac{\dot{u}_2}{W}\,,\qquad
\hat X = \frac{\dot{u}_1}{W}x + \frac{i\hbar}{m} u_1
\frac{\partial}{\partial x}\, .
\label{BasicOperators}
\end{equation}

The new invariant  $\hcI$ plays now the role of $\hat{H}_{HO}$ in this 
more general setting.
The eigenfunctions of this operator, solutions of the Generalized Caldirola--Kanai Schr\"odinger 
equation, are \cite{QAT}:
\begin{eqnarray}\eqalign{
 \phi_\nu(x,t) &=
\frac{1}{\sqrt{\sqrt{2 \pi} \Gamma(\nu+1) \tilde{b}}}
\Bigl(
\frac{\tilde{u}_2 - i \tilde \Omega u_1}
{\tilde{b}}
\Bigr)^{\nu+\frac{1}{2}}
e^{
\frac{i}{2 \hbar} m x^2 \bigl( 
\frac{\tilde\Omega^2 u_1}
{\tilde{u}_2\tilde{b}^2} 
+ \frac{\dot{\tilde{u}}_2}
{\tilde{u}_2 W}\bigr)
}
\\
&
\biggl(
C_1 D_\nu \Bigl(\sqrt{\frac{2 m \tilde\Omega}{\hbar} }\frac{ x}
{\tilde{b}} \Bigr) 
+ 
C_2 D_{-1-\nu} \Bigl(i\sqrt{\frac{2 m \tilde\Omega}{\hbar} }\frac{ x}
{\tilde{b}} \Bigr) 
\biggr)\,,}
\label{autoestadosDMinvariant}
\end{eqnarray}
where $C_1$ and $C_2$ are arbitrary constants,
$\tilde\Omega = \sqrt{\tilde\omega^2- \frac{\tilde\gamma^2}{4}}$ and $\nu$ is
in general a complex number, $D_\nu$ are the parabolic 
cylinder functions  and $\Gamma(z)$ is the Gamma function \cite{Gradshteyn}, 
$u_1$ and $u_2$ are solutions of the LSODE equation corresponding to the given Generalized Caldirola--Kanai oscillator, 
$\tilde{u}_2 = u_2-\tilde \gamma u_1/2$ and  the function
\begin{equation}
\tilde{b}(t)=\sqrt{\tilde{u}_2^2 + \tilde\Omega^2 u_1^2}= \sqrt{(u_2-\tilde \gamma u_1/2)^2 + \tilde\Omega^2 u_1^2}
\end{equation}
plays the role of $b(t)$, i.e. satisfies the Generalized Ermakov equation (\ref{GEP2}) with $\omega_0=\tilde\Omega$, but with different initial conditions, namely
$\tilde{b}(0)=1$ and $\dot{\tilde{b}}(0)=-\frac{\tilde{\gamma}}{2}$.

As in the case without damping, neither $b(t)$ nor $\tilde{b}(t)$ vanish, otherwise the
Wroskian $W(t)$ of the two solutions $u_1$ and $u_2$ would also vanish. 

Thus, the invariant $\hcI$ can be seen as $\hat{H}_{HO}$ mapped from the harmonic 
oscillator with frequency $\tilde\Omega$ through a QAEPT (\ref{Arnold-Ermakov-Pinney2})
characterized by $\tilde{b}(t)$ satisfying (\ref{GEP2}) with $\omega_0=\tilde\Omega$.

The associated spectrum of $\hcI$ is
\begin{equation}
 \lambda_\nu = \hbar \,\tilde\Omega \,(\nu + \frac{1}{2}) \,.
 \label{espectro}
\end{equation}

The integer, real or complex character of $\nu$ depends on the value of $\tilde\Omega$
(and this in turns depends on the particular values of $\tilde \omega$ and $\tilde \gamma$). See \cite{DodonovManko}
for a discussion in the case of the damped harmonic oscillator.

The choice $\tilde{\omega}=\omega_2(0)$ and $\tilde{\gamma}=0$ leads to a generalized Lewis
invariant $\hcIL$ where the arbitrary frequency has been chosen as $\omega_0=\omega_2(0)$,
providing an invariant that commutes with the Generalized Caldirola--Kanai Hamiltonian (\ref{GCK-Hamiltonian}),
and in fact coincides with it,  at $t=0$. This invariant could be useful in short-cuts to 
adiabaticity processes for damped systems or with time-dependent mass (like in waveguides 
\cite{waveguides}), where an invariant commuting with the Hamiltonian is needed.

For the damped harmonic
oscillator with constant $\omega$ and $\gamma$ (see Sec. \ref{SecCK}), there exists a
different choice 
$\tilde \omega=\omega$
and $\tilde \gamma=\gamma$ leading to the only quadratic invariant $\hcI\equiv \hcI_{DM}$, the Dodonov--Man'ko Invariant, 
whose unique, explicit time dependence is through the Wronskian $W(t)$ (like the Caldirola-Kanai Hamiltonian itself). 
Even more, the coherent states associated with this invariant (through a factorization 
of the form $\hcI_{DM}=\frac{1}{2}(\hat{A}^\dag\hat{A}+\hat{A}\hat{A}^\dag)$, $\hat{A}^\dag,\hat{A}$ 
being conserved creation-annihilation operators) are the only ones with minimal, time-independent 
uncertainty relations (see Dodonov \& Man'ko \cite{DodonovManko}).

In conclusion, when studying Generalized Caldirola-Kani systems with the Lewis--Riesenfeld invariant method or with the more general method
given by the QAEPT, a different choice for the
invariant operator than $\hcIL$ should be made, in particular  $\hcI$ 
(with suitable coefficients $\tilde{\omega}$ and $\tilde{\gamma}$) may be
more appropriate (like the case of $\hcIDM$ for the damped harmonic oscillator). We shall denote this 
invariant a Generalized Dodonov--Man'ko  invariant $\hcIGDM$.

\subsection{Engineering a suitable QAEPT to build a Generalized Dodonov--Man'ko invariant}
\label{Engineering}

Once the general setting has been established, let us apply the method in a suitable way to obtain 
a proper invariant $\hcIGDM$.
The previous discussion on the choice of an appropriate invariant, together with the corresponding 
analysis of the Caldirola--Kanai oscillator (with constant damping and frequency, see below),
suggest that it would be helpful to construct a QAEPT from a Generalized Caldirola--Kanai system 2 to a yet 
undetermined Generalized Caldirola--Kanai system 1, but satisfying certain requirements, implemented in 
the choice of $b(t_2)$. In other words: we look for an auxiliary system to help solving a Generalized Caldirola--Kanai
oscillator in such a way that the QAEPT is as simple as possible.

In particular, it is easy to check that choosing $b=W_2^{1/2}$, the Generalized Caldirola--Kanai system 1 is a 
time-dependent 
harmonic oscillator, i.e. it has no damping term. 

In fact, with $b=W_2^{1/2}$ the corresponding QAEPT is given by:
\begin{equation}
 t_1=\int\frac{W_2}{b^2}dt_2=t_2\equiv t\,,\qquad x_1=\frac{x_2}{b}=\frac{x_2}{W_2^{1/2}}\,,
 \label{AEPT-W}
\end{equation}
and
\begin{equation}
 \phi(x_2,t)=W_2^{-1/4} e^{i\frac{m}{4\hbar}\frac{\dot{W}_2}{W_2^2}x_2^2}
\varphi(\frac{x_2}{W_2^{1/2}},t)\,. \label{QAEPT-W}
\end{equation}
The wave function $\varphi$ satisfies the Schr\"odinger equation for a time-dependent 
harmonic oscillator with 
frequency
\begin{equation}
 \omega_1(t)^2=\omega_2(t)^2 +\frac{2 W_2 \ddot{W}_2-3 \dot{W}_2^2}{4 W_2^2}
 =\omega_2(t)^2 - \frac{1}{4}\dot{f}_2^2-\frac{1}{2}\ddot{f}_2\,,
\end{equation}
and the Wronskian for system 1 turns to be $W_1=1$, i.e. the auxiliary system 1 is not damped. However,
$\omega_1(t)$ is not arbitrary but specifically designed to simplify the mapping. Therefore, it is 
straightforward to map results (such as invariants) and computations from the known, auxiliary system 1 to the Generalized Caldirola--Kanai
system 2.

In this case the canonicity conditions (\ref{canonicity-b}) are not satisfied in general, since
$b(0)=1$ but $\dot{b}(0)=\frac{1}{2}\dot{W}(0)$ (see the examples on Sec. \ref{Examples}).

Note that transformation (\ref{AEPT-W}), together with its corresponding extension to velocity, is nothing
other than a generalized version \cite{Pedrosa} of the  time-dependent canonical transformation that removes the damping in the damped harmonic oscillator
\cite{Gyzl} (see also \cite{Expanding1,Expanding2}).

Constructing the Lewis invariant $\hcIL$  for this time-dependent 
harmonic oscillator, and mapping it back to our Generalized Caldirola--Kanai system 2 through
the previous QAEPT, leads to a Generalized Dodonov--Man'ko invariant $\hcIGDM$ for system 2.
More precisely, the invariant can be written as:
\begin{equation}
 \hcIGDM = \frac{\hat{P}^2}{2m}+\frac{1}{2}m\left(\omega_2(0)^2-\frac{1}{2}\ddot{f}_2(0)\right)\hat{X}^2
+\frac{\dot{f}_2(0)}{2} \;\frac{ \hat{X}\hat{P}+\hat{P}\hat{X}}{2}\,.
\label{IGDM}
\end{equation}

We shall provide
examples of this construction in the next section.

\section{Examples}
\label{Examples}

Let us discuss some examples of damped systems, where the previous ideas can be applied to construct invariants
and simplify analytical computations.

\subsection{Caldirola--Kanai oscillator}
\label{SecCK}

The simplest example that can be studied is the quantum damped harmonic oscillator, also known as 
Caldirola--Kanai  oscillator \cite{Caldirola,Kanai}. From the point of view of invariant operators, 
it was first studied by Dodonov \& Man'ko
\cite{DodonovManko}, who constructed first-order invariants in the form of conserved creation
and annihilation operators and derived a basis of \textit{number} states and a family
of coherent states satisfying minimal, time-independent uncertainty relations. The classical equation of motion is given by (we shall not consider the external
force term):

\begin{equation}
\ddot{x}+\gamma \dot{x} +\omega^{2}x=0\,,
\label{CK}
\end{equation}
where $\omega$ and $\gamma$ are constants. 
Even though this is a linear equation with
constant coefficients, the system is not conservative since the Hamiltonian describing this
equation is time-dependent:
\begin{equation}
 H_{CK} = e^{-\gamma t}\frac{p^2}{2m}+\frac{1}{2}m\omega^2 e^{\gamma t}x^2\,.
\end{equation}

There is an old controversy with the quantum version of the Caldirola--Kanai oscillator 
concerning the dissipative character of  this system \cite{Controversy}. 
The main drawback is that the evolution is unitary for all times and there is no loss of 
coherence, something that it is considered inherent to a quantum dissipative system. 
Some proposals have been made to address that paradoxical situation \cite{Dieter} (see also \cite{Dekker}). 

An alternative physical interpretation to the damping term in (\ref{CK}) is that of a 
time-dependent mass, that is, the mass is actually of the form:
\begin{equation}
 m(t)=me^{\gamma t}\qquad\Rightarrow \qquad H_{CK} = \frac{p^2}{2m(t)}+\frac{1}{2}m(t)\omega^2 x^2\,.
\end{equation}

Thus the Caldirola--Kanai oscillator describes an oscillator whose mass is growing exponentially.

The solutions of the classical equations (\ref{CK}), satisfying the canonical conditions $u_1(0)=\dot{u}_2(0)=0,\,u_2(0)=\dot{u}_1(0)=1$ (see \cite{QAT}) are:
\begin{equation}
u_{1}(t)=e^{-\frac{\gamma}{2}t}\, \frac{\sin\Omega t}{\Omega},\qquad
u_{2}(t)=e^{-\frac{\gamma}{2}t}\left(\cos\Omega t + 
\frac{\gamma}{2 \Omega}\sin\Omega t\right),
\label{CK-u1u2}
\end{equation}
where $\Omega=\sqrt{\omega^{2}-\frac{\gamma^{2}}{4}}$ and the Wronskian is given by $W(t) \equiv \dot{u}_{1}(t) u_{2}(t) - u_{1}(t)\dot{u}_{2}(t) =
e^{-\gamma t}=m/m(t)$. The solutions are oscillatory if $\omega>\frac{\gamma}{2}$ (underdamping) and have 
a single zero for  $\omega< \frac{\gamma}{2}$ (overdamping) and $\omega=\frac{\gamma}{2}$ (critical damping).
In this last case the solutions are $u_1(t)=t e^{-\frac{\gamma}{2}t}$ and 
$u_{2}(t)=e^{-\frac{\gamma}{2}t}\left(1 + 
\frac{\gamma}{2}t\right)$.

Let us restrict to the underdamping case. From the previous solutions the Arnold transformation (\ref{CAT}) is given by:
\begin{eqnarray}\eqalign{
\tau& = \frac{u_{1}(t)}{u_{2}(t)}=\frac{\frac{\sin\Omega t}{\Omega}}{\cos\Omega t + 
\frac{\gamma}{2 \Omega}\sin\Omega t} \\
\kappa &=  \frac{x}{u_{2}(t)} =\frac{x}{e^{-\frac{\gamma}{2}t}\left(\cos\Omega t + 
\frac{\gamma}{2 \Omega}\sin\Omega t\right)}\,.}
\end{eqnarray}

The Arnold--Ermakov--Pinney transformation mapping the Caldirola--Kanai oscillator into the standard harmonic oscillator (and its corresponding quantum version) is also easily derived, resulting in:
\begin{eqnarray}\eqalign{
t' &= \int\frac{W(t)}{b(t)^2}dt = \frac{1}{\omega_0}\arctan \omega_0\tau \\
x'&=  \frac{x}{b(t)}\,, } \label{CAEPT-CK}
\end{eqnarray}
where $\omega_0$ is the (arbitrary) frequency of the auxiliary harmonic oscillator, and 
\begin{eqnarray}
 b(t)&=&u_2(t)\sqrt{1+\omega_0^2 \tau^2}=\sqrt{u_2(t)^2+\omega_0^2u_1(t)^2} \nonumber \\
 &=& e^{-\frac{\gamma}{2}t}\sqrt{\left(\cos\Omega t + 
\frac{\gamma}{2 \Omega}\sin\Omega t\right)^2+ \omega_0^2 \frac{\sin^2\Omega t}{\Omega^2}}\,.
\end{eqnarray}
Note that  $b(t)$ never vanishes since this would imply $u_1(t)=u_2(t)=0$ at some time instant,
and thus the Wronskian would also be zero at that time instant, contradicting the fact that $u_1$ and $u_2$ are 
independent solutions of (\ref{CK}). Note also that $b(t)$ satisfies the canonicity conditions (\ref{canonicity-b}).
 
Let us construct an invariant for the  Caldirola--Kanai oscillator. The first possibility
is to construct a generalized Lewis invariant $\hcIL$ with $\tilde\omega=\omega$ and
$\tilde\gamma=0$:
\begin{equation}
 \hcIL = \frac{1}{2m} \hat P^2 + \frac{1}{2} m  \omega^2 \hat X^2\,,
\end{equation}
leading to the invariant constructed by Pedrosa \cite{Pedrosa}.

According to the discussion in 
Sec.~\ref{SecELR}, for a damped system it is more appropriate to use the Dodonov--Man'ko invariant, which 
is built using the solution $u_1(t)$ and the (noncanonical) solution $\tilde{u}_{2}(t)$, 
with $\tilde\omega=\omega$ and $\tilde\gamma=\gamma$:

\begin{equation}
u_{1}(t)=e^{-\frac{\gamma}{2}t}\, \frac{\sin\Omega t}{\Omega} \,,\qquad
\tilde{u}_{2}(t)=u_{2}(t)-\frac{\gamma}{2}u_1(t)= e^{-\frac{\gamma}{2}t}\cos\Omega t\,.
\end{equation}

Therefore $\tilde{b}(t)=\sqrt{\tilde u_2(t)^2+\Omega^2 u_1(t)^2}=e^{-\frac{\gamma}{2}t}$, which also never vanishes.
The Dodonov--Man'ko invariant (\ref{DMinvariant}) for these solutions 
 is given by:
\begin{equation}
\hcIDM=\frac{1}{2m} \hat P^2 + \frac{1}{2} m \omega^2 \hat X^2 +
\frac{\gamma}{2}\; \frac{ \hat{X}\hat{P}+\hat{P}\hat{X}}{2} =
\hat{H}_{CK}+\frac{\gamma}{2}\;\frac{\hat{x}\hat{p}+\hat{p}\hat{x}}{2}\,.
\end{equation}

It can be shown that, for the specific choice $\tilde \omega=\omega$
and $\tilde \gamma=\gamma$ made above, the invariant $\hat{\cI}_{DM}$ coincides with that provided by Nassar \cite{Nassar1,Nassar2}.
It also coincides with the invariant discussed by Cerver\'o \& Villarroel \cite{Cervero-Q}.

The eigenstates of this invariant, solutions of the Caldirola--Kanai Schr\"odinger equation (note that the time-dependent 
phase appearing in the Lewis--Riesenfeld invariant method is included in the QAEPT), are derived from (\ref{autoestadosDMinvariant}):
\begin{equation}
 \phi_n(x,t) =
\frac{1}{\sqrt{\sqrt{2 \pi} 2^n n! }}
e^{-i\bigl(n\Omega + \frac{1}{2}(\Omega+i\frac{\gamma}{2})\bigr) t}
e^{-\frac{m}{2 \hbar} (\Omega+i\frac{\gamma}{2}) e^{\gamma t} x^2}
H_n \Bigl(\sqrt{\frac{m \Omega}{\hbar} } e^{\frac{\gamma}{2}t} x
\Bigr) \,,\label{EstadosDM}
\end{equation}
with eigenvalues $\hbar\Omega(n+\frac{1}{2})$, with $n=0,1,\ldots$ (assuming $\Omega>0$). 
Thus, the result of \cite{DodonovManko} is recovered.

Note that in deriving (\ref{EstadosDM}) it has been crucial to choose the solutions $u_1(t)$ and $\tilde{u}_{2}(t)$,
since in this case $\tilde{b}(t)=e^{-\frac{\gamma}{2}t}=W^{1/2}$ and thus the change of variables in time in (\ref{CAEPT-CK}) 
is trivial, $t'=t$ (as explained in Sec. \ref{Engineering}). Thus, for the case of the damped harmonic oscillator
the Generalized Dodonov-Man'ko invariant $\hcIGDM$ coincide with $\hcIDM$. 
This is a special feature of the damped harmonic oscillator, since $\ddot{f}_2(0)=0$ in this case.

For an arbitrary wave function $\varphi_{HO}$ of the harmonic oscillator, the corresponding wave function
$\phi_{CK}$ in the Caldirola--Kanai system is given by:
\begin{equation}
\phi_{CK}(x,t)=\frac{1}{\sqrt{\tilde{b}}}e^{i\frac{m}{2\hbar}
                 \frac{\dot{\tilde{b}}}{W\tilde{b}} {x}^{2}}{\varphi_{HO}}\Big(\frac{x}{\tilde{b}},t\Big)=
                 e^{\frac{\gamma}{4}t}e^{-i\frac{m}{4\hbar}\gamma e^{\gamma t}{x}^{2}}{\varphi_{HO}}\Big(e^{\frac{\gamma}{2}t}x,t\Big)\,.
\label{QAEPT-CK}         
\end{equation}

This makes apparent the general strategy: make computations in a (simpler) system (e.g. compute $\varphi_{HO}$)
and map them to the system of interest (get $\phi_{CK}$).

It should be stressed that in deriving (\ref{QAEPT-CK}) 
it has not been necessary to solve an eigenvalue equation for an invariant in the Generalized Caldirola--Kanai
variables, (\ref{QAEPT-CK}) is the result of a 
unitary transformation
between two Hilbert spaces. This is a great improvement with respect to the Lewis--Riesenfeld or the Dodonov--Man'ko invariant method, where first
an invariant is found and then its eigenvectors are computed in order to provide a basis of the Hilbert space. With the QAEPT
it is possible to map any harmonic oscillator wave function $\varphi_{HO}$ into its corresponding function $\phi_{CK}$ in the
Caldirola--Kanai Hilbert space. In particular number states, coherent states, squeezed states, or even density matrices 
(see \cite{HarmonicStates}).

\subsection{Hermite oscillator}

Similar considerations can be made in more general systems following the same steps. Let us consider  
a LSODE system that has a
damping rate
linear in time $\gamma=\alpha t$, with $\alpha>0$:
\begin{equation}
\ddot{x}+\alpha t \dot{x} +\omega^{2}x=0\,.
\label{Hermite}
\end{equation}
This equation is similar to the Hermite differential equation, thus this system  is known
as Hermite oscillator \cite{HermiteOscillator}. In analogy with the oscillator with time-dependent mass,
the Hermite oscillator would have a mass
\begin{equation}
 m(t)=m e^{\alpha t^2/2}\,.
\end{equation}

In order to seek for wavefunctions of the quantum Hermite oscillator we need
the solutions of the classical equation (\ref{Hermite}). Those satisfying the canonicity conditions (\ref{canonicity}) are:
\begin{equation}
u_1(t)=t \, _1F_1\left(\frac{1}{2}+\frac{\omega^2}{2 \alpha };\frac{3}{2};-\alpha t^2/2 \right) \,,\quad
u_2(t) = \, _1F_1\left(\frac{\omega^2}{2 \alpha };\frac{1}{2};-\alpha  t^2/2\right)
\end{equation}
and the Wronskian is $W(t)= e^{-\alpha t^2/2}$. Here $_1F_1(a;b;z)$ denotes the confluent hypergeometric function \cite{Gradshteyn}.

From these solutions the Arnold transformation (\ref{CAT}) is given by:
\begin{eqnarray}\eqalign{
\tau& = \frac{u_{1}(t)}{u_{2}(t)}=\frac{t\, _1F_1\left(\frac{1}{2}+\frac{\omega^2}{2 \alpha };\frac{3}{2};-\alpha t^2/2 \right)}{_1F_1\left(\frac{\omega^2}{2 \alpha };\frac{1}{2};-\alpha  t^2/2\right)} \\
\kappa &=  \frac{x}{u_{2}(t)} =\frac{x}{_1F_1\left(\frac{\omega^2}{2 \alpha };\frac{1}{2};-\alpha  t^2/2\right)}\,, }
\end{eqnarray}
and with this the QAT (\ref{QAT}) is easily derived.

The solution $u_2(t)$ has zeros if $\alpha<\omega^2$ (underdamping) and has no zeros if 
$\alpha>\omega^2$ (overdamping) or $\alpha=\omega^2$ (critical damping). In this last case 
$u_2(t)=e^{-\frac{1}{2} t^2 \omega^2}$ and 
$u_1(t)=\frac{1}{\omega}\sqrt{\frac{\pi}{2}} e^{-\frac{1}{2} t^2 \omega^2} \textrm{Erfi}\left(\frac{\omega t}{\sqrt{2}}\right)$. 
Thus, for overdamping and critical damping the Arnolnd transformation is defined for all
times, whereas for the underdamping case the transformation is local in time, mapping a patch of the
Hermite oscillator onto a patch of a free particle trajectory.
Note that in the underdamping case the system performs a finite number of oscillations, since the zeros
of the confluent hypergeometric functions are finite: $u_2(t)$ is even and has 
$2\lceil \frac{\omega^2-\alpha}{2\alpha}\rceil$ zeros  whereas $u_1(t)$ is odd and has 
$2\lceil \frac{\omega^2-2\alpha}{2\alpha}\rceil+1$ zeros.

The Arnold--Ermakov--Pinney transformation (and its corresponding quantum version QAEPT) is also easily derived, resulting in:
\begin{eqnarray}
t' &=& \int\frac{W(t)}{b(t)^2}dt = \frac{1}{\omega_0}\arctan \omega_0\tau \\
x'&=&  \frac{x}{b(t)} 
\end{eqnarray}
where $\omega_0$ is the (arbitrary) frequency of the auxiliary harmonic oscillator, and 
\begin{equation}
 b(t)=u_2(t)\sqrt{1+\omega_0^2 \tau^2}=\sqrt{u_2(t)^2+\omega_0^2u_1(t)^2}
\end{equation}
Note that, as before,  $b(t)$ never vanishes. 

With this the  QAEPT is obtained and the generalized Lewis invariant $\hcIL$ (i.e. the Hamiltonian for the harmonic oscillator mapped into
the Hermite oscillator) can be computed, recovering the results in
 \cite{HermiteOscillator}. Let us give, instead, the construction of a generalized Dodonov--Man'ko invariant as proposed in Sec.
 \ref{Invariant-GCK} and \ref{Engineering}.

Choosing $b(t)=W(t)= e^{-\alpha t^2/2}$ and renaming $x'\equiv y$, the equation (\ref{Hermite}) transforms into
\begin{equation}
\ddot{y} +(\omega^{2} -\frac{\alpha}{2}-\frac{1}{4} \alpha^2 t^2 )y=0\,,
\end{equation}
and the wave functions transform as given in (\ref{QAEPT-W}):
\begin{equation}
 \phi(x,t)=e^{\frac{1}{8}\alpha t^2} 
e^{-i \frac{m}{4\hbar}\alpha  t e^{\frac{\alpha  t^2}{2}}x^2}
\varphi(e^{\frac{1}{4}\alpha t^2}x,t)\,,
\end{equation}
where $\varphi(y,t)$ is a solution of the Schr\"odinger equation for the time-dependent 
harmonic oscillator with 
frequency $\omega^{2} -\frac{\alpha}{2}-\frac{1}{4} \alpha^2 t^2$. Applying the 
Lewis--Riesenfeld invariant method to this time-dependent 
harmonic oscillator we obtain a Lewis invariant that, when mapped to the
original system provides a generalized Dodonov--Man'ko invariant, given by:
\begin{equation}
\hcIGDM=\frac{1}{2m} \hat P^2 + \frac{1}{2} m (\omega^2-\alpha) \hat X^2 
\end{equation}

Note that there is no term in $\frac{ \hat{X}\hat{P}+\hat{P}\hat{X}}{2}$, due to the 
fact that in this case $\dot{f}(0)=0$. This term would appear if we had chosen 
another initial time $t_0\neq 0$.

A similar construction can be performed for the \textit{Lane--Endem} oscillator:

\begin{equation}
\ddot{x}+\frac{\mu}{1+\nu t} \dot{x} +\omega_0^{2}x=0\,,
\label{LaneEndem}
\end{equation}
for which the construction of the Lewis invariant was given in \cite{LaneEmdenOscillator}. We shall only provide here the expression of the Generalized
Dodonov--Man'ko invariant, which turns to be:
\begin{equation}
\hcIGDM=\frac{1}{2m} \hat P^2 + \frac{1}{2} m (\omega^2+\frac{1}{2}\mu\nu) \hat X^2 
+ \frac{\mu}{2}\;\frac{ \hat{X}\hat{P}+\hat{P}\hat{X}}{2}\,.
\end{equation}

\section{Conclusion and outlook}

In this paper the relation between the Lewis-Riesenfneld invariant method and the Quantum Arnold-Ermakov-Pinney transformation
has been established. The former aims at finding an invariant for a time-dependent harmonic oscillator, in order to build a basis of
eigenstates of this invariant satisfying the Schr\"odinger equation (up to a time-dependent phase). The latter is a unitary transformation
that relates two Generalized Caldirola-Kanai systems, and
allows to map states and invariant operators from one system to the other. In particular, if one of the system is a
harmonic oscillator and the other is a time-dependent harmonic oscillator, the Lewis-Riesenfeld invariant method is
recovered. The time-dependent phase is built into the transformation, in such a way that it maps solutions of the Schr\"odinger
equation for one system into solutions for the Schr\"odinger equation of the other directly.

Any invariant of the harmonic oscillator is mapped, through the QAEPT, into an invariant of the time-dependent harmonic
oscillator. In particular, the Hamiltonian for the harmonic oscillator is mapped to the Lewis invariant, explaining why it has discrete eigenvalues.

In this paper we have also shown that the QAEPT is global, in contrast to the QAT, which is local in time. This explains
the robustness and wide applicability of the Lewis-Riesenfeld invariant method.

The main advantage of the QAEPT is that it can also be  applied to damped systems (or with time-dependent mass) in a unified way. In the case
of constant damping and frequency we have shown that a more convenient choice is the Dodonov-Man'ko invariant, rather than the usual Lewis invariant.
The reason is that it shares with the Hamiltonian the loss-energy property \cite{DodonovManko} (which amounts to the only explicit 
time-dependence being through the Wronskian $W(t)$), and therefore its eigenstates or the associated coherent states have nicer properties\footnote{There is an interesting connection between the Dodonov-Man'ko invariant and the
Hamiltonian for the damped harmonic oscillator in expanding coordinates $H_{exp}$, see \cite{Expanding1,Expanding2,DieterNuestro}.}.

We thus propose in the general case to use, as an alternative to the Lewis invariant, the generalized version of the Dodonov-Man'ko invariant when dealing with time-dependent damped 
(Generalized Caldirola-Kanai) systems. This could be useful in designing short-cuts to adiabaticity  processes \cite{InverseEngineer} when
damping or time-dependent masses are present \cite{waveguides,Shortcut-waveguides}, as well as in constructing coherent and related states.


Another interesting application of the QAEPT would be in the case of mixed states (see \cite{HarmonicStates} where  the QAT was discussed in this setting).
Let us denote by $\hat{\rho}_{HO}$ a density matrix for the harmonic oscillator, satisfying the quantum Liouville
equation

\begin{equation}
 \frac{\partial\,\hat{\rho}_{HO}}{\partial t} =- \frac{i}{\hbar} [\hat{H}_{HO},\hat{\rho}_{HO}]\,.
\end{equation}

Then, the density matrix $\hat{\rho}_{GCK}=\hat{E}\hat{\rho}_{HO}\hat{E}^{-1}$ is a proper density matrix
(since $\hat{E}$ is unitary $\textrm{Tr}(\hat{\rho}_{GCK})=\textrm{Tr}(\hat{\rho}_{HO})=1$ ) satisfying the  quantum Liouville equation for the Generalized Caldirola-Kanai
oscillator:

\begin{equation}
\frac{\partial\,\hat{\rho}_{GCK}}{\partial t} =- \frac{i}{\hbar} [\hat{H}_{GCK},\hat{\rho}_{GCK}]\,.
\end{equation}

All properties of the density matrix $\hat{\rho}_{HO}$ are transferred to $\hat{\rho}_{GCK}$, such as characteristic
functions, quasi-probability distributions, etc. Also, since 
$\textrm{Tr}(\hat{\rho}_{GCK}^2)=\textrm{Tr}(\hat{\rho}_{HO}^2)$, the purity or mixed-state character of $\hat{\rho}_{HO}$ is shared
by $\hat{\rho}_{GCK}$. 
In particular, if $\hat{\rho}_{HO}$ describes a  Gaussian state, $\hat{\rho}_{GCK}$ also  corresponds to a Gaussian state.

However, since the QAEPT does not transform  Hamiltonians into each other (only Schr\"odinger and 
quantum Liouville equations do), care should be taken in the physical interpretation of the transformed
density matrix. For instance, a thermal equilibrium state for the harmonic oscillator is not 
mapped into a thermal equilibrium state of the Generalized Caldirola-Kanai
oscillator (in the case when it makes physical sense, for instance when the time scale of the time-dependence of the 
Hamiltonian $\hat{H}_{GCK}$ is much larger than that of relaxation to thermal equilibrium).

A deeper study of the QAEPT applied to mixed states in order to analyze how entanglement is transformed
under $\hat{E}$ (generalized to multipartite systems), 
and how it can be used to describe
dissipation and decoherence analyzing the transformation properties of master equations like
the Lindblad one under $\hat{E}$ is the subject of a work in progress and will be presented
elsewhere.

\section*{Acknowledgments}

Work partially supported by the MCYT, Junta de Andaluc\'\i a and
Fundaci\'on S\'eneca under projects FIS2008-06078-C03-01,
P06-FQM-01951 and 08816/PI/08.

\end{document}